# Collective and separate metal-insulator transitions in correlated vanadium dioxide

*Xuanchi Zhou [1, 2] \*, Xiaohui Yao [1], Wentian Lu [1, 2], Chunwei Yao [1], Xiaomei Qiao [1]*

[1] Key Laboratory of Magnetic Molecules and Magnetic Information Materials of Ministry of Education & School of Materials Science and Engineering, Shanxi Normal University, Taiyuan, 030031, China

[2] Research Institute of Materials Science, Shanxi Key Laboratory of Advanced Magnetic Materials and Devices, Shanxi Normal University, Taiyuan 030031, China

\*Authors to whom correspondence should be addressed: *xuanchizhou@sxnu.edu.cn* (X. Zhou).

**Abstract**

Deciphering the complicated interplay between collective and separate behaviors lies at the heart of first-order metal-insulator transition (MIT) in correlated electron systems, enabling the rational design of exotic electronic states and functionalities. The critical balance between collective and separate behaviors defines a fundamental collective length scale, typically shorter than 5 nm, that governs emergent quantum orders, yet active control over this dichotomy remains elusive. Here, we realize on-demand manipulation on the collective and separate MIT within correlated $VO_2$ system in a reversible fashion. Artificially designing the oxygen deficiency in $VO_2/VO_{2-x}$ homojunction fosters a collective MIT with an extended collective length scale, whereas the introduction of $TiO_2$ interlayer drives a crossover from this collective to a two-step separate MIT via decoupling electronic order parameter. Incorporating mobile hydrogens into the $VO_2/TiO_2/VO_{2-x}$ trilayer enables a reversible control over electronic phase modulations, transitioning two-step MIT towards either one-step MIT or collective electron localization. Such the ionic control over electronic band structure of $VO_2$ flexibly triggers multi-state MIT, a process governed by hydrogen-related band filling. Our findings transform collective length scale from a passive threshold into a dynamic design parameter, establishing a viable handle for engineering collective and separate MIT for adaptive correlated electronics.

**Key words**: Correlated electron system, Collective phenomena, Metal-insulator transition, Ionic evolution, Vanadium dioxide;

**Introduction**

The intricate competition between collective and separate behaviors underpins the fundamental physics in correlated electron systems, giving rise to exotic physical functionality and phenomena ranging from superconductivity [1-3] and first-order phase transition [4-6] to ferroics.[7, 8] The electron correlations in correlated oxides, coupled with the electron-lattice interactions, drive a collective metal-insulator transition (MIT) behavior marked by sharp transition from localized to itinerant electronic states.[9] Notable examples encompass $VO_2$, $Re$NiO$_3$ and $V_2O_3$ systems, which exemplify the irreplaceable role of first-order MIT property in advancing multidisciplinary device applications in correlated electronics,[10, 11] neuromorphic learning [12, 13] and energy conversions.[14-16] One focal point for accessing novel electronic states and functionality in correlated oxides featuring the MIT property lies in understanding the dynamic interplay between collective and separate behaviors.[17] As a representative case, the artificially designed $VO_2/VO_{2-x}$ homojunction via defect engineering enables a collective electronic phase transition decoupled from separate structural transformation, fostering emergent non-equilibrium monoclinic metallic state.[18] Beyond that, the decoupling of electronic and structural phase transformations can be further realized in $VO_2$ system through photoexcitation [11, 19, 20] or isovalent $Ti^{4+}$ doping.[21] This breakthrough not only addresses a long-standing controversy over the physical origin governing the MIT of $VO_2$, but also holds great promise for high-speed correlated electronic devices by virtue of isostructural phase transition.

In particular, imparting a local modulation to an electron-correlated system within collective length scale can trigger collective responses, as evidenced by carrier delocalization of $VO_2$ achieved via electrostatic surface charge accumulation using ionic liquid gating.[22] Nevertheless, next-generation electronics leveraging collective behavior is fundamentally bottlenecked by limited collective length scale in correlated systems, which typically fall below 5 nm.[23, 24] The collective length scale in correlated oxides is dictated by the critical balance between the free energy penalty for sustaining unfavorable collective state and the interfacial energy cost of phase coexistence; below this scale, individual constituents act in unison to form ordered states. To achieve a longer collective length, an alternative approach is to effectively reduce the energy cost associated with the new interface formed between two distinct phases in correlated system.[17, 25] Nevertheless, to date, the ability to facilely tailor the crossover between collective and separate behaviors in correlated oxides remains a formidable challenge, hindering the rational design of emergent functionalities. Consequently, to bridge local perturbations and collective responses in correlated systems, a paradigm shift is imperative: moving beyond passive observation of collective length scales toward active engineering the collective-to-separate crossover in a reversible and controllable fashion.

Correlated $VO_2$ system, featured by a collective first-order MIT property at ~340 K,[26-28] is herein selected as a model system for realizing the facile control over the collective and separate behaviors, unlocking tunable electronic states and functionalities. Harnessing the exceptional electronic band structure of $VO_2$ to oxygen deficiency, the delicate design of $VO_2/VO_{2-x}$ homojunctions enables a collective electronic phase transition, featuring an extended collective length of ~10 nm attributed to similar atomic configurations between constituent phases. Nevertheless, the insertion of an electronically insulating $TiO_2$ layer triggers a crossover in correlated $VO_2$ system from this collective behavior to a two-step separate MIT. Of particular note is the tunable two-step separate MIT realized in $VO_2/TiO_2/VO_{2-x}$ trilayer through protonation, which can transition toward either a one-step collective MIT or complete electronic localization. Utilizing theoretical calculations and advanced spectroscopic techniques, underlying physical picture associated with hydrogen-related band-filling control is unambiguously unraveled. Our findings provide fundamentally new insights into the collective and separate MIT in electron-correlated systems, establishing a cornerstone for the rational design of exotic electronic states leveraging the MIT.

**Results**
**Collective and separate behaviors in correlated electron system**
One critical point for facilely tuning the first-order MIT in correlated $VO_2$ system hinges on deciphering the intricate interplay between collective and separate behaviors, a cornerstone for omnipresent phase coexistence and the design of emergent electronic states (Supplementary Note 1). Capitalizing on the carrier delocalization of VO$_2$ induced by oxygen deficiency via band-filling control, the delicate design of defect engineering offers a fertile ground to tailor the collective and separate behaviors in correlated VO$_2$ system. Introducing the oxygen vacancies into the lattice of VO$_2$ readily donates electron carriers into unoccupied conduction band states, reducing the resultant transition temperature ($T_{MIT}$) (Figure 1a).[29, 30] Collective electronic phase transition is expected for VO$_2$/VO$_{2-x}$ homojunction, as previously reported,[18] wherein both VO$_2$ and oxygen-deficient VO$_{2-x}$ constituents act in unison to showcase a one-step collective MIT (Figure 1b). Beyond accelerated oxygen desorption through vacuum annealing, a built-in gradient for oxygen chemical potential between TiO$_2$ and VO$_2$ allows for directional oxygen ionic migration from VO$_2$ to TiO$_2$, giving rise to an oxygen deficiency in VO$_2$ underlayer.[31, 32] Inserting an electronically insulating TiO$_2$ layer into the VO$_2$/VO$_{2-x}$ system not only introduces oxygen vacancies into the VO$_2$ underlayer, but also drives a separate electronic phase transition via disentangling coupled electronic order parameters (Figure 1c). Consequently, the artificial design of VO$_2$/VO$_{2-x}$ system through decoupling intertwined electronic order parameters via an electronically insulating layer is poised to offer a tunable route for flexibly switching between collective and separate phase transition behaviors in correlated VO$_2$ systems.

**Realizing the crossover between collective and separate MIT in VO$_2$**

To realize the above critical concept, two distinct VO$_2$/VO$_{2-x}$ bilayer and VO$_2$/TiO$_2$/VO$_{2-x}$ trilayer are deposited on *c*-plane Al$_2$O$_3$ template, where the symmetry mismatch between rutile VO$_2$ film and hexagonal Al$_2$O$_3$ substrate results in vertically aligned domain boundaries, a highway for ionic migration.[33, 34] This interpretation is further visualized by respective high-resolution transmission electron microscopy (HRTEM) in Figures 1d and S1, in which the vertically aligned domain boundary in VO$_2$ film induced by Al$_2$O$_3$ template serves as an unobstructed conduit for oxygen ionic evolution. In addition, the domain-matching epitaxy between VO$_{2-x}$ underlayer and Al$_2$O$_3$ substrate is affirmed by HRTEM technique in Figure 1e, in which the expected rutile lattice (for VO$_2$ or VO$_{2-x}$) and hexagonal lattice (for Al$_2$O$_3$) are clearly identified by using Fast Fourier Transform (FFT). The film thicknesses of individual layers in the VO$_2$/TiO$_2$/VO$_{2-x}$ trilayer were determined by using the HRTEM and atomic force microscopy (AFM) techniques (Figures S2-S3) to be approximately 20 nm (VO$_2$ overlayer), 2 nm (TiO$_2$ interlayer), and 10 nm (VO$_{2-x}$ underlayer). In addition, the surface roughness of as-fabricated VO$_2$/TiO$_2$/VO$_{2-x}$ system is also characterized by AFM technique, revealing a root-mean-square roughness below 0.5 nm (Figure S4). The introduction of oxygen vacancies results in the lattice expansion along the *out-of-plane* direction in comparison with pristine counterpart, as exemplified by the leftward shift of VO$_2$ (002) diffraction peak in their X-ray diffraction (XRD) patterns (Figure S5). Two distinct diffraction peaks representing VO$_2$ and VO$_{2-x}$ are clearly resolved in the XRD pattern of VO$_2$/TiO$_2$/VO$_{2-x}$ trilayer.

In order to characterize atomic-scale elementary distribution, energy dispersion spectrum (EDS) analysis was conducted on the VO$_2$/TiO$_2$/VO$_{2-x}$ trilayer, with the results displayed in Figures 1f and S6. The relatively sharp interface is observed for the inserted $TiO_2$ interlayer with a thickness of around 2 nm in the $VO_2/TiO_2/VO_{2-x}$ trilayer, free of obvious elementary intermixing. Further consistency in elemental spatial distribution is verified by using time-of-flight secondary ion mass spectrometry (ToF-SIMS) analysis, where the molecular ions sputtered from the $VO_2/TiO_2/VO_{2-x}$ trilayer using an incident ion beam (e.g., $Cs^+$) are effectively detected by the mass spectrometer (Figure 1g). The insertion of $TiO_2$ interlayer in $VO_2$ system can be directly identified by three-dimensional ToF-SIMS element maps, which align well with the depth profiles of titanium concentration, where the titanium signal appears between $VO_2$ overlayer and $VO_{2-x}$ underlayer over sputtering time (Figure 1h). The reduction in the vanadium signal from the $VO_{2-x}$ underlayer relative to the $VO_2$ overlayer may arise from signal attenuation caused by electronically insulating $TiO_2$ interlayer. Together, these observations validate the controlled design of a $VO_2/TiO_2/VO_{2-x}$ heterostructure comprising an ultrathin $TiO_2$ interlayer that effectively separates the $VO_2$ overlayer and $VO_{2-x}$ underlayer with distinct and chemically isolated interfaces, free from pronounced elementary intermixing.

The flexible switching between collective and separate phase transitions in correlated $VO_2$ systems can be flexibly controlled by just inserting an electronically insulating $TiO_2$ interlayer, as demonstrated by their temperature dependences of material resistivity ($\rho$-$T$) in Figure 2a. Incorporating oxygen vacancies into the $VO_2$ underlayer, as driven by either vacuum annealing ($VO_2/VO_{2-x}$ bilayer) or chemical potential mismatch ($VO_2/TiO_2/VO_{2-x}$ trilayer), reduces the resulting $T_{MIT}$ of $VO_2$ (Figure S7). This understanding is demonstrated by using Raman technique in Figure 2b, in which the characteristic Raman peaks (e.g., $\omega_1$ and $\omega_2$ peaks) deriving from the V-V dimerization are depressed via oxygen vacancies, a typical signature of carrier delocalization.[5, 35] The $VO_2/VO_{2-x}$ bilayer, in which each constituent exhibits a distinct $T_{MIT}$, showcases a collective MIT behavior akin to that of $VO_2$, as further demonstrated by respective temperature coefficient of resistance (*TCR*)-*T* tendency in Figure 2c. Nevertheless, the incorporation of an electronically insulating $TiO_2$ interlayer decouples intertwined electronic order parameters in $VO_2$ system, enabling the crossover from a collective MIT to a two-step separate MIT that behaves as individual constituents of $VO_2$ and $VO_{2-x}$ (Figure S8). Consequently, the inserted $TiO_2$ interlayer acts as an oxygen reservoir that accommodates oxygen ions transferred from the underlying $VO_2$, while also enabling a two-step MIT functionality. Given the electrical shunting effect from the $VO_{2-x}$ underlayer, optimizing the thickness ratio between the $VO_2$ overlayer and the $VO_{2-x}$ underlayer is critical for unveiling genuine two-step separate MIT behavior (Figure S9). In particular, the collective length herein achieved in the $VO_2/VO_{2-x}$ system (e.g., ~10 nm) significantly exceeds the previously reported ones in superconducting coherence length, ferroelectric vortex and skyrmion, a hallmark for collective quantum phenomena (Figure 2d).[36-40] This observation stems from comparable atomic configurations between $VO_2$ and oxygen-deficient $VO_{2-x}$, which effectively reduces the interfacial energy penalty for the newly created boundary during the separate MIT.[17] Such an extended collective length scale not only enables a more robust collective phase transition against local perturbations, but also improves the tunability of electronic states, offering a versatile platform for low-power correlated electronic devices.

**Tunable two-step separate MIT in VO$_2$ through protonation**

Hydrogen doping offers a versatile pathway for tailoring the MIT functionality in VO$_2$ system, transitioning correlated electronic ground state ($t_{2g}^1 e_g^0$) to either electron-itinerant state ($t_{2g}^{1+\Delta} e_g^0$) or electron-localized state ($t_{2g}^2 e_g^0$).[41, 42] Such the two-step insulator-metal-highly insulator transformation pathway realized in VO$_2$ through protonation extends the horizons for designing correlated electronic states and MIT property. The incorporation of hydrogens driven by hydrogen spillover strategy (Figure S10) induces an *out-of-plane* lattice expansion of VO$_2$, as evidenced by the leftward shift of its diffraction peak due to O-H interactions. The low-temperature hydrogenation at 100 ºC for 1 h gives rise to an emergent diffraction peak at 36.6 º corresponding to an insulating hydrogenated phase (Figure 3a). Remarkably, the two-step separate MIT in the VO$_2$/TiO$_2$/VO$_{2-x}$ trilayer can be readily adjusted by using hydrogen doping, driving the system toward either a one-step MIT or an electron-localized state (Figure 3b). Unlike the two-order-of-magnitude enhanced resistivity of the heavily hydrogenated VO$_2$ monolayer relative to the pristine counterpart,[43] the electron-localized VO$_2$/TiO$_2$/VO$_{2-x}$ trilayer shows slower hydrogenation kinetics to induce relative lower material resistivity, owing to its extra heterointerfaces and oxygen deficiency. Performing the hydrogenation at 70 ºC for 30 min substantially depresses the MIT of VO$_2$ overlayer via hydrogen-related electron doping, resulting in a one-step MIT behavior. However, the excessive hydrogenation at 100 ºC for 1-5 h instead results in the collective electron localization of VO$_2$/TiO$_2$/VO$_{2-x}$ trilayer, displaying typical insulating transport behavior. Incorporating hydrogens into VO$_2$ lattice readily drives a crossover from a two-step separate MIT to a one-step MIT, and ultimately to collective electron localization. Elevating the hydrogenation temperature and prolonging the treatment period are expected to introduce more hydrogen into VO$_2$ lattice, triggering sequential electronic phase modulations along the insulator-metal-highly insulator route.

Further consistency in hydrogen-related phase transformations in VO$_2$ system is validated by corresponding Raman spectra in Figure 3c, in which hydrogen doping suppresses the V-V dimerization of VO$_2$, irrespective of whether the system exhibits one-step MIT behavior or electronic localization. The ultrahigh mobility of incorporated hydrogens enables reversible hydrogen-associated electronic phase modulations in this VO$_2$ system through oxidative annealing at identical conditions (e.g., 70 ºC, 30 min) (Figure 3d), as further validated by the *TCR-T* curves (Figure S11). In addition, hydrogenated electronic states realized in VO$_2$/TiO$_2$/VO$_{2-x}$ trilayer is relatively robust via exposure to ambient atmosphere, in contrast to pristine VO$_2$, whose high-resistive state rapidly recovers within 10 h (Figures 3e and S12).[33, 44] This relatively sluggish dehydrogenation kinetics observed for VO$_2$/TiO$_2$/VO$_{2-x}$ trilayer aligns well with the aforementioned hydrogenation process. The robust but reversible electronic phase modulations in correlated $VO_2$ system through hydrogenation, spanning from a two-step separate MIT to a one-step collective MIT and ultimately to electron localization, benefit correlated electronic device applications using multi-state MIT. The incorporation of hydrogen is evidenced by ToF-SIMS technique in Figures 3f-3g, where a significantly higher hydrogen intensity is observed for $VO_2/TiO_2/VO_{2-x}$ trilayer region than for the $Al_2O_3$ substrate. The incorporated hydrogen concentration progressively diminishes away from the hydrogen source located at the film surface. The ionic control over electronic state evolutions in $VO_2$ is dictated by electronic orbital reconfiguration via protonation, wherein partial occupation of the $d_{//}^*$ orbital drives metallization, while integer electron filling of the $d_{//}^*$ orbital through extensive hydrogen incorporation instead leads to the electron localization (Figure 3h).[42, 43]

**Hydrogen-related band-filling control over electronic band structure**

The underlying variations in chemical environment of $VO_2/TiO_2/VO_{2-x}$ system through hydrogenation is corroborated by X-ray photoelectron spectra (XPS) analysis, as the V $2p$ and O $1s$ core-level spectrum shown in Figures 4a-4b. The introduction of hydrogen donates electron carriers into the electronic band structure of $VO_2$, gradually reducing the valence state of vanadium from $V^{4+}$ towards $V^{3+}$ (Figure 4a). Upon hydrogenation, an elevated intensity in the O-H interaction (e.g., ~532 eV) is observed for $VO_2$ with respect to the V-O interaction (e.g., ~529.5 eV), consistent with ToF-SIMS result (Figure 4b). Synchrotron-based element-specific soft X-ray absorption spectra (sXAS) analysis provides specific evidences for hydrogen-induced orbital reconfiguration in $VO_2$ system (Figures 4c-4d). The V-$L$ edge spectrum of $VO_2$ deriving from the V $2p \rightarrow 3d$ transition serves as an effective indicator for the changes in the vanadium valence states,[45] the leftwards shift of which unravels the reduction in the oxidation state (Figure 4c). Notably, the orbital hybridization between V-$3d$ and O-$2p$ orbitals, along with the presence of intrinsically empty O-$2p$ states in $VO_2$, allows for the feasible detection of the electron filling in V-$3d$ orbital of $VO_2$ through proton evolution (Figure 4d).[46] Hydrogenation leads to a reduced relative intensity of the first peak of O $1s$ spectrum representing the low-energy $t_{2g}$ band, compared with the second peak that arises from $e_g$ band,[47] which aligns well with that of $VO_2/VO_{2-x}$ bilayer via defect-mediated band filling (Figure S13). This finding indicate that the doped electron carriers as introduced by protonation readily occupy the low-energy $t_{2g}$ orbital of $VO_2$. Such hydrogen-related variations in valence states and band structure of $VO_2$ are further amplified by prolonging the hydrogenation period, underscoring the correlation between hydrogen concentration and phase modulations.

The physical origin of hydrogen-related electronic phase modulations in $VO_2$ system is further investigated by using density functional theory (DFT) calculations via analyzing the electronic density of states (DOS) (Figures 4e-4g). Pristine $VO_2$ exhibits a band gap of ~0.8 eV (Figure 4e), consistent with its correlated electron ground state as reported in prior study.[9] Remarkably, hydrogen intercalation into the VO$_2$ lattice (e.g., H$_7$V$_8$O$_{16}$) induces a metallic phase characterized by a finite DOS at the Fermi level, signifying the emergence of an electron-itinerant state. Instead, increasing the hydrogen concentration to the stoichiometric levels (e.g., one H per vanadium atom) restores an insulating state, with an obvious band gap being detected. Combined with sXAS result, hydrogenation initially induces orbital reconfiguration of VO$_2$ towards an electron-itinerant state based on $t_{2g}^{1+\Delta}e_g^0$ configuration through the partial filling in the $t_{2g}$ band. Nevertheless, the near-integer occupation in the $t_{2g}$ band of VO$_2$ drives the formation of electron-localized state via opening a new band gap between $d_{//}^*$ orbital and $\pi^*$ orbital, as also confirmed by calculated band structure (Figures S14-S15). Benefiting from hydrogen-related band-filling control, two-step separate MIT realized in VO$_2$ system via inserting an electronically insulating interlayer can be reversibly adjusted towards either one-step collective MIT or collective electron localization.

**Discussion**

The complicated interplay between collective and separate behaviors underpins a cornerstone for understanding the first-order MIT in an electron-correlated system. The collective length, a critical scale below which constituents act in unison, has traditionally served only as a passive boundary: collective MIT dominates below it, separate transitions above it. Extending this length is therefore essential to stabilize collective behavior over larger spatial scales, enabling robust correlated functionalities. Similar atomic configuration between VO$_2$ and oxygen-deficient VO$_{2-x}$ allows for an extended collective length scale through depressing interfacial energy cost. In particular, the ability to actively toggle between collective and separate MIT in correlated system transforms a static physical threshold into a dynamic design parameter, unlocking unprecedented control over emergent electronic states. Incorporating mobile ionic species (e.g., oxygen vacancies and hydrogen) meanwhile offers a viable and controllable pathway to achieve this dynamic tunability between collective and separate MIT in correlated oxides through band-filling control. Such the active control is expected to enable multistate memory, neuromorphic plasticity, and reconfigurable logic within a single material platform, overcoming the inherent rigidity of first-order MIT and opening a new frontier for adaptive electronics.

In summary, we demonstrate a facile handle to achieve facile switching between collective and separate MIT in correlated VO$_2$ system through the incorporation of mobile ionic species, engendering controllable multi-state MIT. Analogous lattice framework between VO$_2$ and oxygen-deficient VO$_{2-x}$ enables an enlarged collective length of ~10 nm in VO$_2$/VO$_{2-x}$ system, in which individual layers with distinct $T_{MIT}$ showcase a collective one-step MIT. Nevertheless, the insertion of electronically insulating TiO$_2$ interlayer results in a two-step separate MIT via decoupling intertwined electronic order parameters. Leveraging hydrogen-related band-filling control, such the two-step MIT realized in VO$_2$/TiO$_2$/VO$_{2-x}$ trilayer can be reversibly adjusted, transitioning towards either one-step MIT or collective electronic localization. Spectroscopic techniques and theoretical calculations reveal that hydrogen-driven electronic phase modulations in $VO_2$ are governed by the electron filling in the $t_{2g}$ band, which drives multi-state electronic orbital reconfiguration. Our present work establishes controllable ionic evolution as a direct and reversible pathway to switch between collective and separate behaviors in correlated system, enabling a programmable multi-state MIT for adaptive correlated electronics.

**Methods**

**Sample preparation:** The $VO_2$ heterostructures were deposited on the single crystalline *c*-plane $Al_2O_3$ substrates through the laser molecular beam epitaxy (LMBE) technique. The deposition temperature, the oxygen partial pressure, the target-substrate distance, and the laser fluence for depositing $VO_2$ underlayer were optimized as 500 °C, 1.5 Pa, 45 mm, and 1.0 J cm$^2$, respectively. Following the growth of underlying $VO_2$ layer, the substrate temperature was lowered to 250 °C and the oxygen pressure was reduced to 1 Pa for the subsequent deposition of the $TiO_2$ layer. Finally, the $VO_2$ overlayer was grown at a temperature of 300 °C under an oxygen pressure of 1.5 Pa. Afterwards, the $VO_2/TiO_2/VO_2$ heterostructures were cooled down to room temperature at a rate of 20 °C per minute under an identical oxygen partial pressure. Oxygen deficiency is introduced into the lattice of $VO_2$ through either high-vacuum annealing or chemical potential mismatch between $VO_2$ and $TiO_2$. To realize an effective hydrogenation, 20 nm-thick platinum dots were sputtered into the surface of as-fabricated $VO_2$ heterostructures via magnetron sputtering, while $Pt/VO_2/TiO_2/VO_2$ heterostructures were afterward annealed in a 5 % $H_2$/Ar forming gas.

**Characterization:** The crystal structures of the $VO_2/TiO_2/VO_{2-x}$ heterostructures were characterized by using the X-ray diffraction (XRD) (Rigaku, Ultima IV) and high-resolution transmission electron microscopy (HRTEM) (FEI, Tecnai G2 20 S-TWIN). Raman spectra (HORIBA, HR Evolution) was exploited to probe the material microstructure of the heterostructure. The chemical environment and valence state for the $VO_2/TiO_2/VO_{2-x}$ heterostructures were characterized by using the X-ray photoelectron spectroscopy (XPS) technique (Thermo, K-Alpha X). The surface roughness and film thickness of the grown $VO_2/TiO_2/VO_{2-x}$ heterostructures are identified by using atomic force microscope (AFM) (Bruker, Dimesion Icon). The electronic structure of $VO_2/TiO_2/VO_{2-x}$ heterostructures was explored via using soft X-ray absorption spectroscopy (sXAS) analysis, which was conducted at the Shanghai Synchrotron Radiation Facility (SSRF) on beamline BL08U1A and National Synchrotron Radiation Laboratory (NSRL) on beamline BL12B-b. Electrical transport properties for the $VO_2/TiO_2/VO_{2-x}$ heterostructures were further probed using a commercial physical property measurement system (PPMS) (Quantum design), while the room-temperature material resistance were measured by using a Keithley 4200 system. The elementary depth profile is examined by using the time-of-flight secondary ion mass spectrometry (ToF-SIMS) technique (ION-TOF GmbH, TOF.SIMS 5).

**First-principles calculations:** First-principles calculations were performed using the projector augmented wave (PAW) method [48] implemented in the QUANTUM ESPRESSO package. A 2×1×1 supercell expansion along the *a*-axis was employed to construct the hydrogenated $VO_2$ ($H_7V_8O_{16}$ and $H_8V_8O_{16}$), with different hydrogen atoms incorporated per unit cell. The calculations utilized a plane-wave energy cutoff of 90 Ry and a 5×11×9 Monkhorst-Pack k-point mesh for Brillouin zone sampling. Electron correlation effects in V-3*d* orbitals were addressed through the generalized gradient approximation (GGA) + U approach,[49] with a Hubbard U parameter of 3.32 eV applied to V-3*d* states. Structural optimization was conducted by fully relaxing atomic coordinates until inter-atomic forces fell below $10^{-3}$ Ry/Bohr. Self-consistent field calculations employed an energy convergence threshold of $10^{-12}$ Ry to ensure high numerical precision. Hydrogen-modified electronic band structure was subsequently calculated using the optimized geometry, enabling an detailed analysis of hydrogen-induced electronic state modifications.

**Data availability**

The data that supports the findings of this study are available from the corresponding authors upon reasonable request.

**Acknowledgements**

This work was supported by the National Natural Science Foundation of China (No. 52401240), Fundamental Research Program of Shanxi Province (No. 202403021212123), and Scientific and Technological Innovation Programs of Higher Education Institutions in Shanxi (No. 2024L145). The authors acknowledge the beam line BL08U1A at the Shanghai Synchrotron Radiation Facility (SSRF) (https://cstr.cn/31124.02.SSRF.BL08U1A) and the beam line BL12B-b at the National

**Author contributions**

X.Z. conceived this study, and lead the project; X.Z. planned for the experiment, and analyzed the results; X.Y., and X.Q. grew $VO_2$ films, and carried out the transport measurements under the supervision of X.Z.; W.L. performed the DFT calculations assisted by C.Y.; X.Z. wrote the paper with contributions from all authors; All authors discussed the results and commented on the final manuscript.

**Competing interests**

The authors declare no competing interests.

**Additional information**

Supplementary information is available for this paper.

**Correspondences:** Correspondences should be addressed: Prof. Xuanchi Zhou (*xuanchizhou@sxnu.edu.cn*).

# Figure Legends

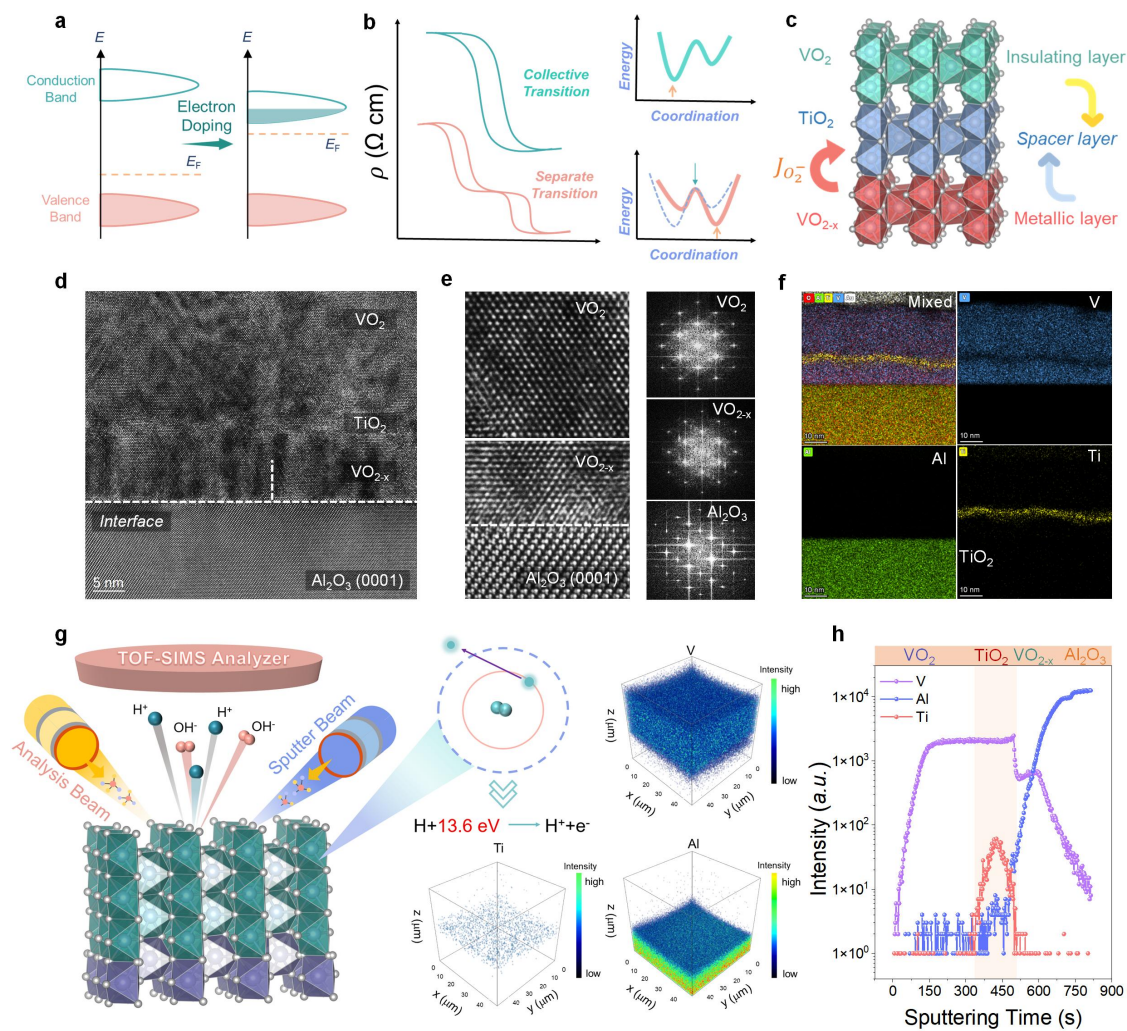

**Fig. 1 | Collective and separate metal-insulator transition (MIT) in correlated VO$_2$. a**, Schematic of band-filling control over electronic band structure of correlated electronic system. **b**, Collective and separate MIT properties in correlated system. **c**, Realizing the collective and separate MIT in VO$_2$/TiO$_2$/VO$_{2-x}$ system through inserting an electronically insulating TiO$_2$. **d**, High-resolution transmission electron microscopy (HRTEM) images for VO$_2$/TiO$_2$/VO$_{2-x}$ trilayer. **e**, Zoom-in images for cross-sectional HRTEM and respective Fast Fourier Transform (FFT) images for VO$_2$/TiO$_2$/VO$_{2-x}$ trilayer. **f**, Vanadium, titanium and aluminum elementary distributions for as-grown VO$_2$/TiO$_2$/VO$_{2-x}$ trilayer characterized by using energy dispersion spectrum (EDS). **g**, Schematic of the principle in time-of-flight secondary ion mass spectrometry (ToF-SIMS) and three-dimensional ToF-SIMS element maps for VO$_2$/TiO$_2$/VO$_{2-x}$ trilayer. **h**, Depth-profile of the elementary distribution in VO$_2$/TiO$_2$/VO$_{2-x}$ trilayer.

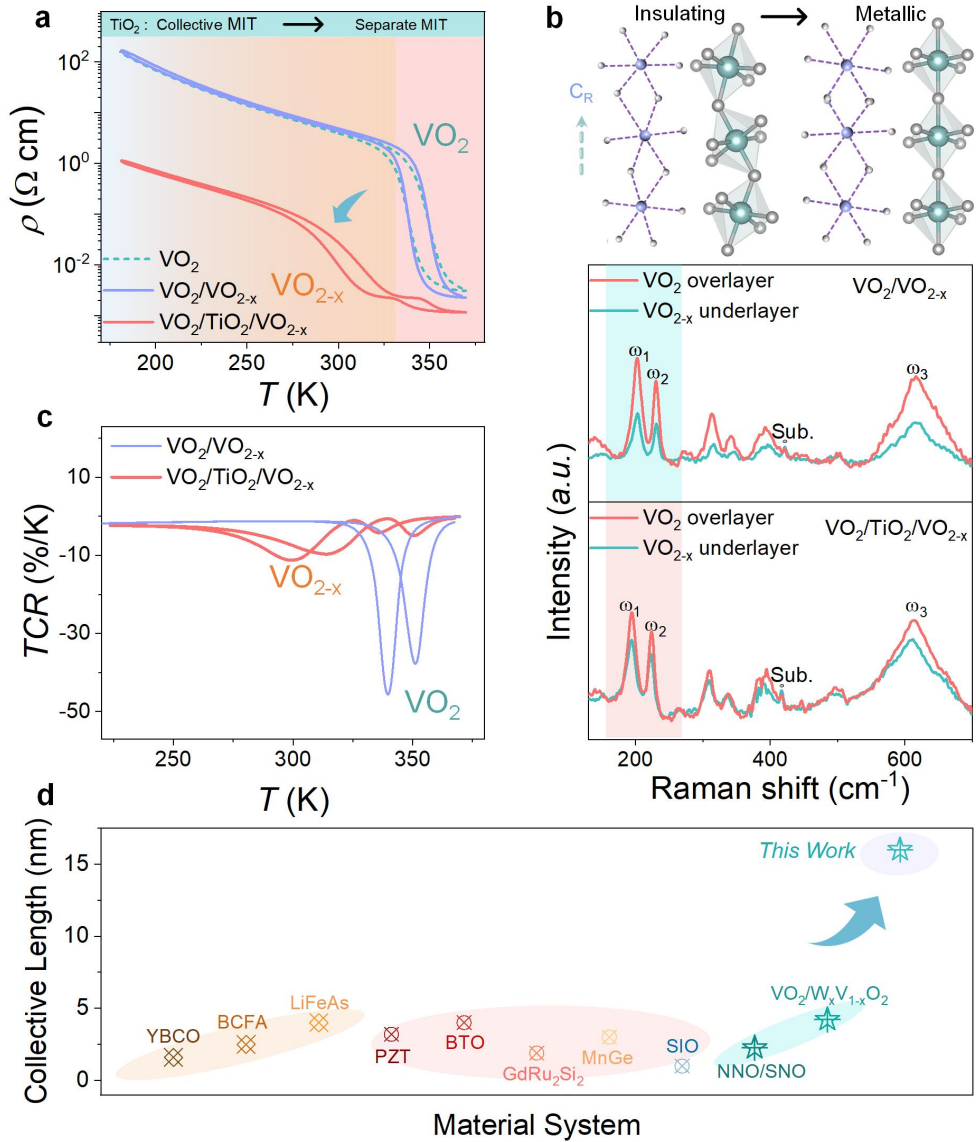

**Fig. 2 | The facile switching between collective and separative MIT in VO₂. a**, Temperature dependence of material resistivity ($\rho$-$T$) as compared for VO$_2$/VO$_{2-x}$ bilayer and VO$_2$/TiO$_2$/VO$_{2-x}$ trilayer. **b**, Raman spectra compared for VO$_2$ and oxygen-deficient VO$_{2-x}$ in VO$_2$/VO$_{2-x}$ bilayer and VO$_2$/TiO$_2$/VO$_{2-x}$ trilayer. **c**, Temperature coefficient of material resistivity compared for VO$_2$ monolayer, VO$_2$/VO$_{2-x}$ bilayer and VO$_2$/TiO$_2$/VO$_{2-x}$ trilayer. **d**, The collective length scale achievable in VO$_2$/VO$_{2-x}$ system, in comparison with other quantum material systems, covering superconducting, ferroic and phase change materials.[36-40]

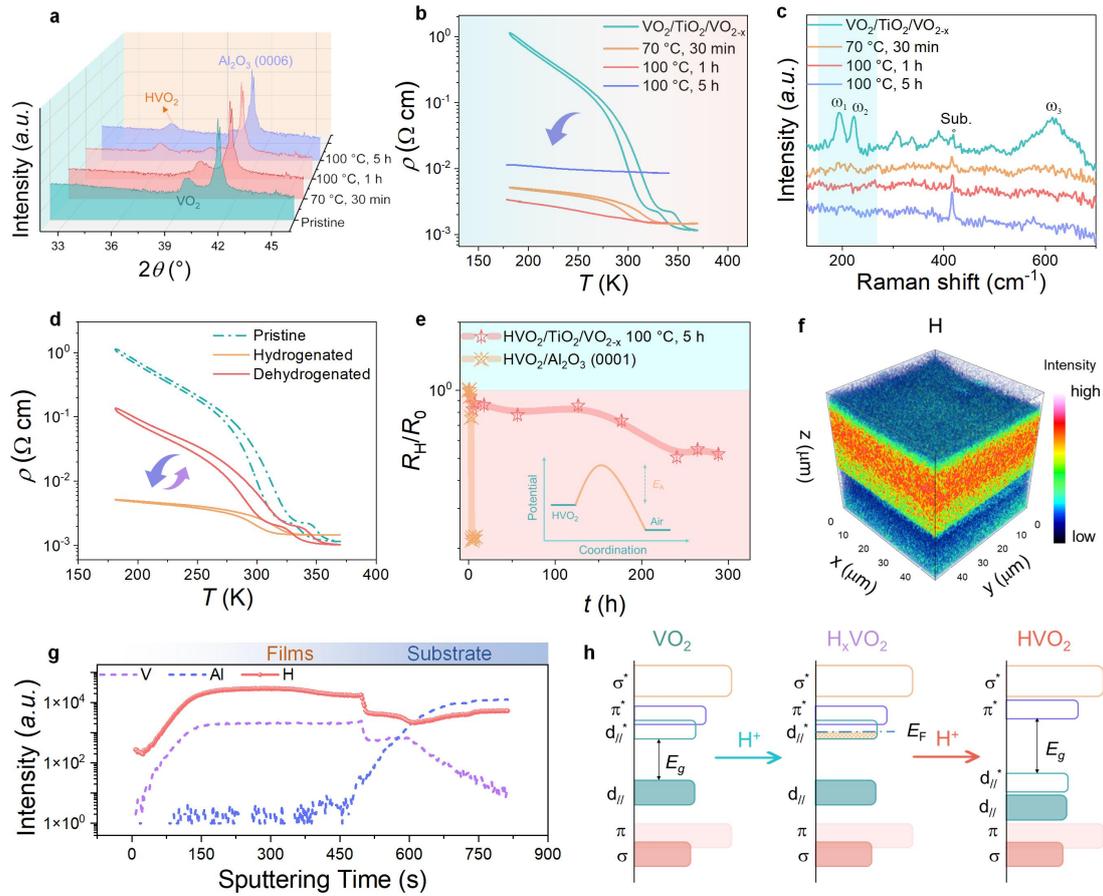

**Fig. 3 | Tunable two-step MIT in VO₂ system through hydrogenation. a**, Comparing the X-ray diffraction (XRD) patterns for $VO_2/TiO_2/VO_{2-x}$ trilayer through hydrogenation. **b**, $\rho$-$T$ tendency compared for hydrogenated $VO_2/TiO_2/VO_{2-x}$ system. **c**, Raman spectra for $VO_2/TiO_2/VO_{2-x}$ system via hydrogenation. **d**, Reversible electronic phase modulations in $VO_2/TiO_2/VO_{2-x}$ system via hydrogenation. **e**, Temporal evolution for hydrogen-induced variation in material resistivity ($R_H/R_0$) in hydrogenated $VO_2/TiO_2/VO_{2-x}$ system. **f**, Three-dimensional ToF-SIMS hydrogen element maps for hydrogenated $VO_2/TiO_2/VO_{2-x}$ trilayer. **g**, Depth-profile of the hydrogen elementary distribution in $VO_2/TiO_2/VO_{2-x}$ trilayer. **h**, Schematic of hydrogen-triggered electronic orbital reconfiguration in $VO_2$.

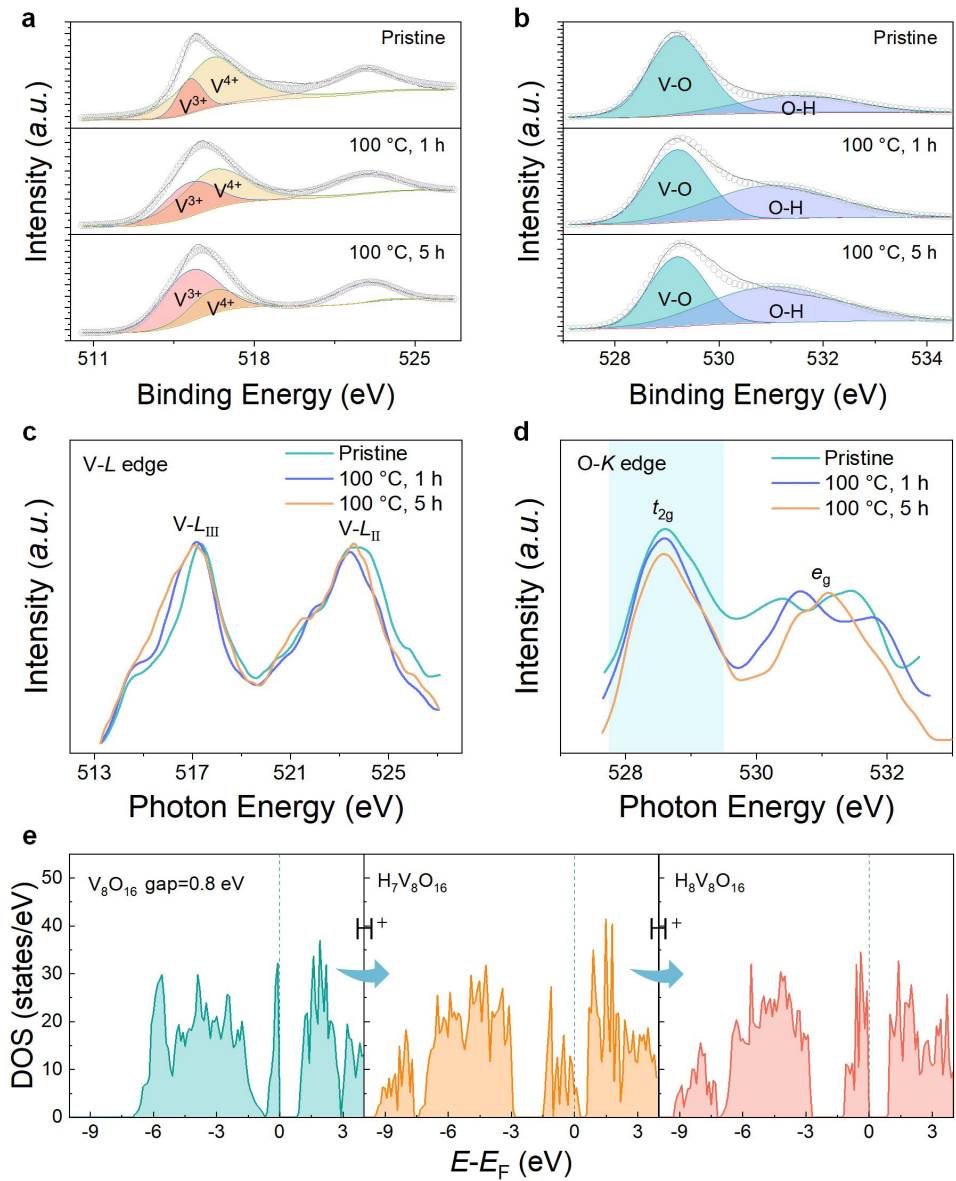

**Fig. 4 | Hydrogen-triggered electronic orbital reconfiguration in VO$_2$. a-b**, X-ray photoelectron spectra (XPS) spectra as compared for the **a**, vanadium and **b**, oxygen core levels of VO$_2$/TiO$_2$/VO$_{2-x}$ system under different hydrogenation conditions. **c-d**, **c**, V-$L$ edge and **d**, O-$K$ edge of soft X-ray absorption spectra (sXAS) for VO$_2$/TiO$_2$/VO$_{2-x}$ through hydrogenation. **e**, Calculated density of states (DOS) of VO$_2$ film **c**, before and after hydrogenation. **d-e**, Calculated fat band structure for VO$_2$/TiO$_2$/VO$_{2-x}$ system through hydrogenation.